%% file: main.tex
\documentclass[11pt]{article}

\usepackage[margin=1in]{geometry}
\usepackage{amsmath,amssymb,amsthm,booktabs}
\usepackage{graphicx}
\usepackage{hyperref}
\usepackage[numbers,sort&compress]{natbib}
\newtheorem{proposition}{Proposition}

\newcommand{\methodname}{PMM-MASEM}
\newcommand{\plugin}{Plugin\_Estimator}
\newcommand{\kensemble}{k\_Ensemble}
\newcommand{\mleexp}{MLE-Exp}
\newcommand{\pmmest}{PMM2/MLE}

\title{Variance-Reduced Manifold Sampling via Polynomial-Maximization Density Estimation}
\author{
Serhii Zabolotnii\thanks{ORCID: 0000-0003-0242-2234. Corresponding author:
\href{mailto:zabolotnii.serhii@csbc.edu.ua}{zabolotnii.serhii@csbc.edu.ua}.}\\
\small Department of Information, Multimedia Technologies and Design,\\
\small Cherkasy State Business College, Cherkasy, 18028, Ukraine\\
\small State Scientific Research Institute of Armament and Military Equipment Testing and Certification,\\
\small Cherkasy, Ukraine\\
\small Department of Cybernetics and Applied Mathematics,\\
\small Uzhhorod National University, Uzhhorod, Ukraine
}
\date{Preprint, 2026-05-24}

\begin{document}
\maketitle

\input{sections/00_abstract}
\input{sections/01_introduction}
\input{sections/02_related_work}
\input{sections/03_theoretical_framework}
\input{sections/04_method}
\input{sections/05_experiments}
\input{sections/06_results}
\input{sections/07_discussion}
\input{sections/08_conclusion}

\section*{Data and Code Availability}

The code supplement, unit tests, experiment drivers, generated CSV outputs,
tables, and figures used to support this manuscript are available in the
\href{https://github.com/SZabolotnii/Ku-PMM-MASEM-code-supplement}{public
GitHub repository}.  No restricted third-party dataset is used; all
experiments are synthetic or simulation-based.

\appendix
\input{appendix/appendix}

\bibliographystyle{plainnat}
\bibliography{refs}

\end{document}

%% file: sections/00_abstract.tex
\begin{abstract}
Uniform sampling on implicitly defined manifolds is a core primitive in motion
planning, constrained simulation, and probabilistic machine learning.  MASEM
addresses this problem by entropy-maximizing resampling, but its
resampling weights depend on a local $k$-nearest-neighbour density estimate
whose errors can be amplified by aggressive resampling temperatures.  We
ask whether a polynomial-maximization moment estimator can replace the
plug-in density rule without changing the surrounding MASEM architecture.
The proposed \methodname{} module computes shell spacings from nested
$k$-nearest-neighbour radii, estimates their standardized cumulants, and uses
a gated PMM2 estimator only when the spacing distribution departs from the
flat $\mathrm{Exp}(1)$ regime.  Symmetric platykurtic panels are treated by a
guarded PMM3-location extension, evaluated separately from the production
PMM2/MLE selector.
This fallback is essential: on a flat homogeneous manifold the plug-in
estimator is already the MLE, so PMM should not outperform it.  A local
Known-DGP Monte Carlo experiment confirms this gate: the selector returns MLE on flat
$\mathrm{Exp}(1)$ spacings, reduces density MSE by 22--36\% on asymmetric gamma
and boundary-spacing regimes, and shows that PMM3-location can reduce mean and
density MSE for strongly platykurtic positive spacings when strict symmetry,
kurtosis, variance-coefficient, and Newton-root checks are met.  A lightweight
resampling-proxy experiment improves seven-lobes coverage but degrades the sine
and swiss-roll proxies.  The current evidence therefore supports an
applicability-boundary result rather than a general MASEM improvement claim.
\end{abstract}

\noindent\textbf{Keywords:} density estimation; manifold sampling; resampling;
$k$-nearest neighbours; simulation; reproducibility.

%% file: sections/01_introduction.tex
\section{Introduction}
\label{sec:introduction}

Uniform sampling on constrained manifolds appears whenever a system is
specified by equalities and inequalities rather than by an explicit
parametrization.  Typical examples include feasible robot configurations,
implicit generative constraints, molecular or mechanical systems with
conservation laws, and posterior slices in probabilistic models.  The
difficulty is not only geometric.  A sampler must represent all connected
components of the feasible set and redistribute mass toward underrepresented
regions without collapsing onto the components that happen to be easiest to
reach from the initialization.

Manifold Sampling via Entropy Maximization (MASEM) \citep{braun2026masem}
addresses this problem by alternating local manifold-constrained rejuvenation
with an entropy-based resampling step.  If $x_i$ denotes a particle and
$\varepsilon_{i,k}$ its distance to the $k$-th nearest neighbour, MASEM uses a
weight of the form
\[
  w_i \propto \varepsilon_{i,k}^{\tau},
\]
equivalently $w_i \propto \hat q(x_i)^{-\tau}$ for the plug-in density
estimate
\[
  \hat q(x_i)=\frac{k}{N V_p\varepsilon_{i,k}^p}.
\]
This rule gives larger weights to particles in low-density regions and is the
mechanism by which MASEM spreads mass across disconnected components.  It is
also the statistical bottleneck.  Braun et al. explicitly note that larger
temperatures ``can amplify errors in the k-NN density estimate''
\citep{braun2026masem}; their practical mitigation is to ensemble several
neighbourhood sizes, which they report as lower-variance but biased.

We ask whether that density-estimation subroutine can be replaced by
a more principled moment-based estimator while preserving the rest of MASEM.
The candidate is the polynomial maximization method (PMM), a family of closed
form estimators for non-Gaussian errors developed in the Kunchenko school of
moment estimation \citep{kunchenko1992stochastic,warsza2018pmm}.  In its PMM2 form, the
standard variance coefficient is
\begin{equation}
  \label{eq:pmm2_variance_coefficient}
  g_2 = 1-\frac{c_3^2}{2+c_4}.
\end{equation}
The coefficient in Eq.~\eqref{eq:pmm2_variance_coefficient} relates estimator
variance to the standardized third and fourth cumulants $(c_3,c_4)$ of the
error distribution.  The MASEM density problem is not a
direct regression problem, so this coefficient cannot be imported blindly.
Instead, we use it as a diagnostic and design principle for a density module
based on the shell spacings
\[
  \Delta_{i,j}=V_p\left(\varepsilon_{i,j}^p-\varepsilon_{i,j-1}^p\right).
\]

A central constraint is the flat case.  On a flat homogeneous manifold, the
normalized shell spacings are asymptotically i.i.d. $\mathrm{Exp}(1)$ and the
plug-in estimator is exactly the MLE.  In that regime PMM has no statistical
room to improve the density estimate.  Our implementation therefore treats
the flat $\mathrm{Exp}(1)$ cumulant signature $(c_3,c_4)=(2,6)$ as an explicit
fallback condition.  The proposed PMM2/MLE rule is activated only when the
observed spacing distribution shows curvature or boundary misspecification.

\begin{figure}[t]
\centering
\includegraphics[width=0.92\linewidth]{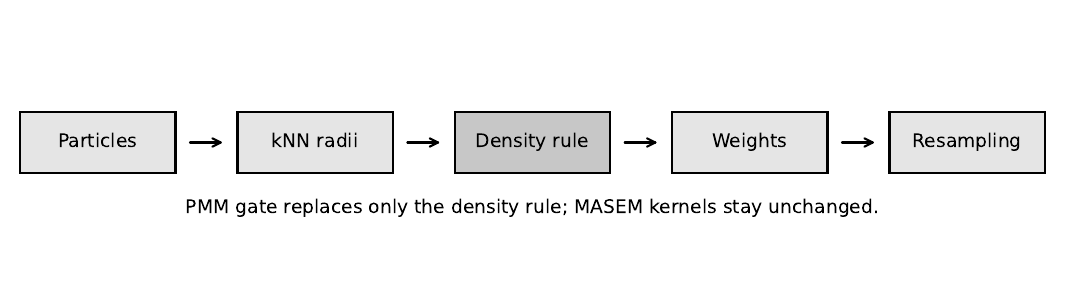}
\caption{PMM-MASEM pipeline.  The PMM gate replaces only the density rule used
for resampling; the surrounding MASEM rejuvenation architecture is unchanged.}
\label{fig:pipeline}
\end{figure}

The contributions are:
\begin{enumerate}
  \item a PMM-gated replacement for the MASEM density weight that preserves the
    NHR/OLLA rejuvenation architecture and changes only the resampling density
    estimate;
  \item an explicit applicability boundary showing why flat homogeneous
    manifolds must fall back to the plug-in/MLE estimator;
  \item a verification protocol and local evidence showing when PMM2 helps,
    when the MLE fallback is statistically required, and when a guarded
    PMM3-location extension is plausible for symmetric platykurtic spacings;
  \item a reproducibility plan with fixed seeds, generated tables and figures,
    and an AI-use disclosure for the final manuscript bundle.
\end{enumerate}

The resulting evidence is deliberately mixed: the gate preserves the flat MLE
case and improves local density MSE in asymmetric spacing regimes; the
PMM3-location study identifies a narrower platykurtic regime where mean and
density MSE can decrease under strict guards; and the proxy experiments expose
failure modes that preclude a general MASEM-superiority claim.

Before describing the method, we position the work relative to manifold
sampling, nonparametric density estimation, and PMM moment estimators
(\S\ref{sec:related_work}).

%% file: sections/02_related_work.tex
\section{Related Work}
\label{sec:related_work}

\paragraph{Manifold sampling and entropy resampling.}
Sampling on implicitly defined manifolds has been approached through
rejection methods, constrained Markov chains, sequential Monte Carlo, and
local tangent-space kernels.  MASEM \citep{braun2026masem} is the direct
starting point for this paper because it addresses disconnected feasible sets
by combining local feasible rejuvenation with entropy-driven resampling.  Its
resampling step uses an empirical density surrogate to move probability mass
towards underrepresented regions.  The present work does not replace the
MASEM architecture, local kernels, or constraint handling: it isolates the
statistical role of the density estimate inside the resampling weights.

\paragraph{$k$-nearest-neighbour density estimation.}
The MASEM plug-in rule descends from classical multivariate $k$-nearest
neighbour density estimation \citep{loftsgaarden1965nonparametric}.  Mack and
Rosenblatt \citep{mack1979multivariate} analysed its bias--variance behaviour
and emphasized the dependence on $k$, tail behaviour, and the local
approximation of density by a ball volume.  Recent analyses sharpen this
picture by separating risk criteria and support assumptions: for example,
Zhao and Lai \citep{zhao2020analysis} show that boundary knowledge can change
whether $k$NN density estimation is minimax-optimal under strong uniform
criteria.  This literature sets the baseline for our method: a single-radius
plug-in estimator is strong in the locally Euclidean case, whereas boundary or
support misspecification can create a real opportunity for a guarded
replacement.

\paragraph{Neighbour distances, spacings, and the Poisson null model.}
Nearest-neighbour distances also underlie spacing and entropy estimators.
The Kozachenko--Leonenko entropy estimator \citep{kozachenko1987sample}
uses $k$NN radii to estimate local log-density, while the classical theory of
spacings \citep{pyke1965spacings} and Poisson processes
\citep{kingman1993poisson,penrose2003random} gives the reference law for
shell-volume increments.  In a locally homogeneous flat model, the increments
of nested ball volumes behave as independent exponential spacings after the
local intensity is factored out.  This Poisson case is the null model of the
present paper: under $\mathrm{Exp}(1)$ spacings the plug-in density rule is
also the MLE, so any PMM module must fall back rather than claim improvement.

\paragraph{Density estimation on manifolds and near boundaries.}
Non-Euclidean density estimation makes the preceding local-ball approximation
more delicate.  Pelletier \citep{pelletier2005kernel} adapts kernel density
estimation to compact Riemannian manifolds using geodesic structure and volume
corrections; Henry, Mu{\~n}oz, and Rodriguez \citep{henry2011nearest} study
$k$NN-type density estimation when observations belong to a Riemannian
manifold.  Boundary effects form a separate source of misspecification: Berry
and Sauer \citep{berry2017density} develop density estimators for manifolds
with boundary by estimating boundary direction and distance.  These works
motivate our experimental regimes.  Curvature and boundary truncation are not
incidental nuisances for MASEM; they are precisely the mechanisms that can move
shell-spacing cumulants away from the flat exponential reference law.

\paragraph{Moment estimators and PMM.}
The polynomial maximization method (PMM) provides moment-based corrections for
non-Gaussian error regimes through standardized cumulants
\citep{kunchenko1992stochastic,warsza2018pmm}.  The PMM2 coefficient
in Eq.~\eqref{eq:pmm2_variance_coefficient} is attractive here because the
shell-spacing distribution can be summarized by $(c_3,c_4)$ before a density
rule is chosen.  Recent
preprints and software work extend this estimator family to non-Gaussian
regression and time-series settings, including the EstemPMM implementation and
a large PMM2--ARIMA simulation study
\citep{zabolotnii2026estempmm,zabolotnii2025arimapmm2}.  We use these works as
documentation that PMM2/PMM3 form an implementable cumulant-based estimation
family, not as direct evidence for the MASEM density task.  That task is not a
standard linear regression, measurement, or time-series problem.  The
present paper is therefore an applicability study, not a direct
transplantation of PMM theory: when the spacing law is $\mathrm{Exp}(1)$ the
plug-in/MLE rule is retained; when curvature or boundary effects move the
cumulants away from that regime, PMM2 becomes a guarded candidate; PMM3
motivates a separate location-equation study for symmetric platykurtic spacing
panels.  That PMM3-location branch is interpreted as conditional evidence for
positive spacings, not as a replacement for the production PMM2/MLE selector.
Adaptive fractional bases and PATP-style selectors are left to a separate
follow-up study rather than included in the present evidence package.

These observations motivate a theoretical analysis of $k$-NN shell spacings on
flat and curved manifolds (\S\ref{sec:theory}).

%% file: sections/03_theoretical_framework.tex
\section{Theoretical Framework}
\label{sec:theory}

This section isolates the statistical object modified by \methodname{}: the
local density estimate inside MASEM's resampling step.  The surrounding
manifold kernel is not changed.

\subsection{Flat Homogeneous Manifold}

Let $\mathcal M$ be a $p$-dimensional flat manifold and let $\rho$ be locally
constant at the scale of the $k$-NN ball around $x_i$.  Define shell spacings
\[
  \Delta_{i,j}=V_p(\varepsilon_{i,j}^p-\varepsilon_{i,j-1}^p),
  \qquad \varepsilon_{i,0}=0.
\]
Under the standard Poisson approximation for local neighbour counts, the
normalized spacings
\[
  s_{i,j}=N\rho(x_i)\Delta_{i,j}
\]
are asymptotically i.i.d. $\mathrm{Exp}(1)$ as $N\to\infty$ with fixed $k$.
Consequently, $\mathbb E[s_{i,j}]=1$, $\mathrm{Var}(s_{i,j})=1$, and the
standardized cumulants are $c_3=2$, $c_4=6$.

\begin{proposition}[Plug-in estimator is the flat-case MLE]
Under the flat homogeneous assumptions, the MLE for $\rho(x_i)$ based on the
shell spacings is
\[
  \hat\rho_{\mathrm{MLE}}(x_i)
  =\frac{k}{N\sum_{j=1}^k\Delta_{i,j}}
  =\frac{k}{N V_p\varepsilon_{i,k}^p},
\]
which is exactly the MASEM plug-in estimator.
\end{proposition}

The first consequence is methodological: a PMM replacement should not be applied
in this regime.  Any claimed improvement over the plug-in estimator on a truly
flat homogeneous manifold would be inconsistent with the likelihood solution
under the correct model.  Our algorithm therefore uses
the empirical cumulant signature $(c_3,c_4)\approx(2,6)$ as a fallback gate.

\subsection{Curvature and Boundary Misspecification}

On a Riemannian manifold the Euclidean volume term $V_p r^p$ is only a local
approximation to the geodesic ball volume.  For a constant-curvature patch,
\[
  \mathrm{vol}_g(B(x,r))
  =
  V_pr^p\left(1-\frac{\kappa r^2}{6(p+2)}+O(\kappa^2 r^4)\right),
\]
with boundary terms adding further non-Euclidean corrections.  The plug-in
estimator still uses the flat $V_pr^p$ volume, so it becomes an MLE for a
misspecified model.  The resulting spacing distribution need not be
$\mathrm{Exp}(1)$; its cumulants shift away from $(2,6)$ in a manner governed
by curvature, boundary truncation, density variation, $k$, and $N$.

Accordingly, the main text uses a conservative qualitative claim:
PMM is admissible only when the measured shell-spacing law departs from the
flat signature.  The exact leading-order formulas in
\path{theory/riemannian_manifold.tex} require a separate audit before they should
be promoted as theorem-level claims.  The experiments in
\S\ref{sec:experiments} are designed to validate or falsify those formulas on
known-DGP patches.

\subsection{From Density Error to Component Loss}

MASEM uses weights $w_i\propto\hat\rho(x_i)^{-\tau}$.  If
$\hat\rho(x_i)=\rho(x_i)(1+\varepsilon_i)$, then
\[
  \hat\rho(x_i)^{-\tau}
  \approx
  \rho(x_i)^{-\tau}
  \left(1-\tau\varepsilon_i+
  \frac{\tau(\tau+1)}{2}\varepsilon_i^2\right).
\]
Thus density-estimation variance is magnified by $\tau^2$ in the weight
variance.  This is the technical reason PMM, if it reduces density error in a
misspecified curved regime, could allow a larger safe temperature.

For a component $\mathcal C_c$ with effective mass floor
$(\Phi\alpha)_c$, the usual occupancy argument gives a component-loss bound of
the form
\[
  P_c \leq \exp\{-N(\Phi\alpha)_c\}.
\]
The role of the density estimator is indirect: lower weight noise increases
the range of $\tau$ for which the effective floor remains reliable.  The final
paper reports this as a conditional bridge, not as a new convergence
theorem, unless the proof is fully completed.

For the symmetric platykurtic PMM3 diagnostic we use the efficiency coefficient
\begin{equation}
  \label{eq:pmm3_variance_coefficient}
  g_3 = 1-\frac{\gamma_4^2}{6+9\gamma_4+\gamma_6},
\end{equation}
where $\gamma_4$ and $\gamma_6$ are the fourth and sixth cumulants of the
centered symmetric residual law.  This coefficient is used only to screen
whether PMM3 has statistical room in symmetric spacing regimes.

\input{tables/tab1.tex}

This analysis leads to the PMM module and its
integration into MASEM.

%% file: tables/tab1.tex
\begin{table}[t]
\centering
\caption{Empirical kNN-spacing regimes in the Known-DGP Monte Carlo.  Values are means over five seeds; $g_2$ is the coefficient in Eq.~\eqref{eq:pmm2_variance_coefficient}, and $g_3$ is the PMM3 efficiency coefficient for symmetric distributions in Eq.~\eqref{eq:pmm3_variance_coefficient}.  Neither coefficient is interpreted as an end-to-end gain.}
\label{tab:spacing_regimes}
\resizebox{\linewidth}{!}{%
\begin{tabular}{lrrrrrr}
\toprule
Regime & $c_3$ & $c_4$ & $g_2$ & $\gamma_6$ & $g_3$ & selector \\
\midrule
Flat Exp(1) & 1.962 & 5.624 & 0.495 & 125.768 & 0.825 & MLE fallback \\
Mild curved gamma & 2.410 & 8.841 & 0.460 & 359.536 & 0.802 & PMM2 \\
Strong curved gamma & 3.010 & 12.981 & 0.394 & 506.897 & 0.727 & PMM2 \\
Boundary mixture & 2.272 & 8.442 & 0.494 & 426.228 & 0.825 & PMM2 \\
Platykurtic uniform & 0.004 & -1.198 & 1.000 & 6.835 & 0.302 & PMM3-location diagnostic \\
Platykurtic beta & 0.001 & -0.865 & 1.000 & 3.891 & 0.645 & PMM3-location diagnostic \\
\bottomrule
\end{tabular}%
}
\end{table}

%% file: sections/04_method.tex
\section{Method}
\label{sec:method}

\methodname{} changes one part of MASEM: the density estimate used to compute
resampling weights.  Initialization, local manifold rejuvenation, NHR/OLLA
kernels, and slack handling remain unchanged.

\subsection{Baseline MASEM Weight}

For particles $x_1,\ldots,x_N$ on a $p$-dimensional manifold, baseline MASEM
computes the $k$-NN radius $\varepsilon_{i,k}$ and uses
\[
  \hat\rho_{\mathrm{Plugin}}(x_i)
  =
  \frac{k}{N V_p\varepsilon_{i,k}^p},
  \qquad
  w_i=
  \frac{\hat\rho_{\mathrm{Plugin}}(x_i)^{-\tau}}
       {\sum_{\ell=1}^N\hat\rho_{\mathrm{Plugin}}(x_\ell)^{-\tau}}.
\]
The PMM module replaces only $\hat\rho_{\mathrm{Plugin}}$.

\subsection{Shell Spacings and Cumulants}

For each particle we compute sorted radii
$\varepsilon_{i,1}\le\cdots\le\varepsilon_{i,k}$ and shell spacings
\[
  \Delta_{i,j}=V_p(\varepsilon_{i,j}^p-\varepsilon_{i,j-1}^p).
\]
The normalized spacings are pooled across particles to estimate standardized
cumulants $(c_3,c_4)$.  Pooling is necessary because a single particle gives
only $k$ spacings, which are too few for stable cumulant estimation.  The
current implementation uses global pooling; component-aware or local pooling
is an experimental extension.

\subsection{Selector}

The production selector has two density gates and one separate diagnostic gate
for PMM3-location.

\paragraph{Flat gate.}
If $(c_3,c_4)$ is close to $(2,6)$, the module returns the MLE/plug-in density.
This gate enforces the flat-manifold applicability boundary.

\paragraph{PMM2 gate.}
If the spacing law is not Exp(1)-like and the skewness gate is active, the
module computes
\[
  \hat\rho_{\mathrm{PMM2}}(x_i)
  =
  \frac{1}{N\bar\Delta_i}
  \left(1+\frac{c_3(m_{1i}-m_1)}{2+c_4}\right),
\]
where $\bar\Delta_i=k^{-1}\sum_j\Delta_{i,j}$, $m_{1i}$ is the per-particle
mean of normalized spacings, and $m_1$ is the pooled mean.

\paragraph{PMM3 gate.}
For symmetric platykurtic panels we evaluate a PMM3-location candidate rather
than folding it into the main density selector unconditionally.  Let
$\mu_i=\mathbb E[\Delta_{i,\cdot}]$ denote the positive spacing location for a
particle.  The centered PMM3 score is applied to residuals
$\Delta_{i,j}-\mu_i$ and solved for $\mu_i$ by Newton iteration.  The candidate
is accepted only when the pooled diagnostics satisfy
$|c_3|\le0.3$, $c_4<-0.7$, the PMM3 coefficient in
Eq.~\eqref{eq:pmm3_variance_coefficient} satisfies $0<g_3<0.8$, $k\ge16$, and
the Newton root remains positive inside a trust region.  Otherwise the density
for that panel is the MLE value $1/(N\bar\Delta_i)$.  This keeps the production
selector conservative while still testing whether PMM3 has statistical room in
symmetric platykurtic spacing regimes.

\subsection{Algorithm}

\begin{center}
\begin{minipage}{0.96\linewidth}
\hrule
\vspace{0.35em}
\textbf{Algorithm 1: PMM2/MLE density module with PMM3-location diagnostic}
\vspace{0.35em}
\hrule
\vspace{0.45em}
\small
\textbf{Input:} particles $x_{1:N}$, intrinsic dimension $p$, neighbourhood size $k$, temperature $\tau$.\\
\textbf{Output:} normalized production weights $w_{1:N}$ and optional PMM3-location diagnostics.\\[0.25em]
\begin{tabular}{@{}r@{\quad}p{0.86\linewidth}@{}}
\textbf{1.} & For each particle $x_i$, compute sorted radii
  $\varepsilon_{i,1}\le\cdots\le\varepsilon_{i,k}$.\\
\textbf{2.} & Convert radii to shell spacings
  $\Delta_{i,j}=V_p(\varepsilon_{i,j}^{p}-\varepsilon_{i,j-1}^{p})$.\\
\textbf{3.} & Normalize spacings $s_{i,j}=N\Delta_{i,j}$ and pool them over all $i,j$.\\
\textbf{4.} & Estimate pooled standardized cumulants $(c_3,c_4)$ from $s_{i,j}$.\\
\textbf{5.} & Compute the fallback density
  $\hat\rho_{\mathrm{MLE},i}=1/(N\bar\Delta_i)$.\\
\textbf{6.} & \textbf{if} $(c_3,c_4)$ is Exp(1)-like, i.e. close to $(2,6)$, \textbf{then}\\
& \hspace*{1.2em}set $\hat\rho_i\leftarrow\hat\rho_{\mathrm{MLE},i}$.\\
\textbf{7.} & \textbf{else if} the PMM2 denominator and variance checks are valid \textbf{then}\\
& \hspace*{1.2em}set $\hat\rho_i\leftarrow\hat\rho_{\mathrm{PMM2},i}$.\\
\textbf{8.} & \textbf{else if} $|c_3|$ is small, $c_4<-0.7$, $0<g_3<0.8$, $k\ge16$, and the PMM3-location root is valid \textbf{then}\\
& \hspace*{1.2em}record the diagnostic candidate $\hat\rho_{\mathrm{PMM3loc},i}$ and keep the production density at $\hat\rho_{\mathrm{MLE},i}$.\\
\textbf{9.} & \textbf{else}\\
& \hspace*{1.2em}set $\hat\rho_i\leftarrow\hat\rho_{\mathrm{MLE},i}$.\\
\textbf{10.} & Return
  $w_i=\hat\rho_i^{-\tau}/\sum_{\ell=1}^{N}\hat\rho_\ell^{-\tau}$.
\end{tabular}
\vspace{0.35em}
\hrule
\end{minipage}
\end{center}

\subsection{Implementation Status}

The repository currently implements the selector in
\path{src/masem/pmm_module.py}, the estimator registry in
\path{src/masem/estimators.py}, and MASEM integration in
\path{src/masem/masem.py}.  Unit tests validate shape, normalization, JAX tracing,
flat fallback behaviour, PMM2 activation on asymmetric spacing laws, and
PMM3 cumulant diagnostics on symmetric platykurtic spacing laws.  Separate
research drivers evaluate PMM3-location for regression residuals and positive
spacing panels; these diagnostics are reported as conditional evidence rather
than as a production MASEM claim.

\S\ref{sec:experiments} verifies the theoretical predictions on synthetic and
benchmark manifolds.

%% file: sections/05_experiments.tex
\section{Experiments}
\label{sec:experiments}

The experiments are designed to answer two questions separately.  First, does
the PMM selector improve local density estimation when the spacing law is
known to deviate from $\mathrm{Exp}(1)$?  Second, does any local improvement
translate into end-to-end MASEM sampling quality?

\subsection{Experimental Setup}

The four estimators are fixed throughout:
\[
  \plugin,\quad \kensemble,\quad \mleexp,\quad \pmmest.
\]
We evaluate two empirical layers.  The first layer is a Known-DGP density
Monte Carlo in
\path{experiments/run_known_dgp_mc.py}.  It generates flat
$\mathrm{Exp}(1)$, asymmetric gamma, boundary-mixture, and platykurtic spacing
laws with known target density.  Only this layer is used to support a local
density-estimation claim.

The second layer is a lightweight resampling proxy in
\path{experiments/run_resampling_proxy.py}.  It uses biased particle clouds
on disconnected disks, seven lobes, a sine curve, a swiss roll, a scaling
stress test, and a robotics-corridor proxy, then applies the four density
weight rules to a matched resampling step.  This layer tests whether the
weight rule has a plausible operational effect, but it is not a full MASEM
NHR/OLLA benchmark.  We therefore treat it as operational diagnostic evidence
rather than as an end-to-end superiority benchmark.

The third layer is a PMM3 platykurtic diagnostic study.  The regression script
\path{experiments/research_pmm3_platykurtic_regression.R} checks the standard
centered PMM3 location estimator on symmetric distributions with increasing
negative kurtosis.  The spacing script
\path{experiments/research_pmm3_platykurtic_spacing.py} then applies the same
score logic to positive shell-spacing panels through a guarded location solve.
This layer is used only to delimit the symmetric platykurtic case; it does not
replace the full MASEM benchmark.

Each experiment uses five seeds and reports mean $\pm$ 95\% confidence
intervals.  Pairwise significance tests use Holm--Bonferroni correction with
threshold $p<0.01$.  Metrics are:
\begin{itemize}
  \item Wasserstein-2 squared distance (Sinkhorn approximation), the primary
    end-to-end metric;
  \item KL divergence between pairwise-distance histograms;
  \item maximum slack and feasibility diagnostics;
  \item component-loss frequency;
  \item wall-clock overhead.
\end{itemize}

\subsection{Known-DGP Density Microbenchmark}

Known-DGP patches isolate density estimation from MASEM dynamics.  The flat
patch validates the fallback rule: Exp(1)-like spacings should select
Plugin/MLE.  Curved and boundary patches test whether PMM2/MLE reduces density
MSE relative to Plugin and $k$-ensemble baselines.  The generated output is
\texttt{results/known\_dgp\_mc.csv}.

\subsection{Resampling-Proxy Sampling}

The proxy experiments compare the four density rules under matched particle
budgets, $k$, $\tau$, and synthetic target distributions.  They report
nearest-neighbour $W_2^2$ proxy, pairwise-distance histogram KL, component
loss, and estimator wall-clock.  Since no local manifold rejuvenation is run,
the slack metric is fixed at zero and the results are interpreted only as
weight-rule diagnostics.

\subsection{\texorpdfstring{$\tau$-Tolerance}{tau-Tolerance}}

The $\tau$ sweep tests whether lower density-estimation error permits a larger
temperature before W2 or component loss degrades.  This is the operational
meaning of the theory bridge in \S\ref{sec:theory}; the output is
\texttt{results/tau\_tolerance.csv}.

\subsection{Component Coverage}

Component coverage is evaluated by tracking whether each disconnected
component or lobe receives at least one final particle.  For synthetic tasks
with known components, this is direct.  For swiss roll and robotics tasks, the
component/lobe proxy is specified before running the experiment.

\subsection{Computational Overhead}

The PMM module adds cumulant estimation and correction terms to the density
step.  Since all operations are vectorized, the expected overhead is moderate,
but the wall-clock effect is measured empirically.  Proxy wall-clock results are stored in
\texttt{results/wallclock.csv}.

All CSV files, tables, and figures are generated by the repository-wide
experiment driver.  \S\ref{sec:results} summarizes the resulting evidence and
states the remaining end-to-end gap explicitly.

%% file: sections/06_results.tex
\section{Results}
\label{sec:results}

\paragraph{H1: local density estimation.}
The Known-DGP microbenchmark supports the intended flat fallback and gives a
conditional positive result for asymmetric misspecification.  On flat
$\mathrm{Exp}(1)$ spacings, the selector chooses MLE in all five seeds and the
PMM2/MLE MSE ratio against MLE is exactly 1.000.  On asymmetric regimes the
selected PMM rule reduces density MSE: the ratio is 0.688 on mild gamma
spacings, 0.639 on strong gamma spacings, and 0.785 on the boundary-mixture
proxy.  The platykurtic uniform and beta stress cases are handled by the MLE
branch in the production PMM2/MLE selector; their PMM2/MLE ratio is 1.000
across the uniform sensitivity sweep.  Thus H1 is confirmed for the guarded
PMM2 asymmetric regimes and for the flat MLE boundary.

\begin{figure}[t]
\centering
\includegraphics[width=0.95\linewidth]{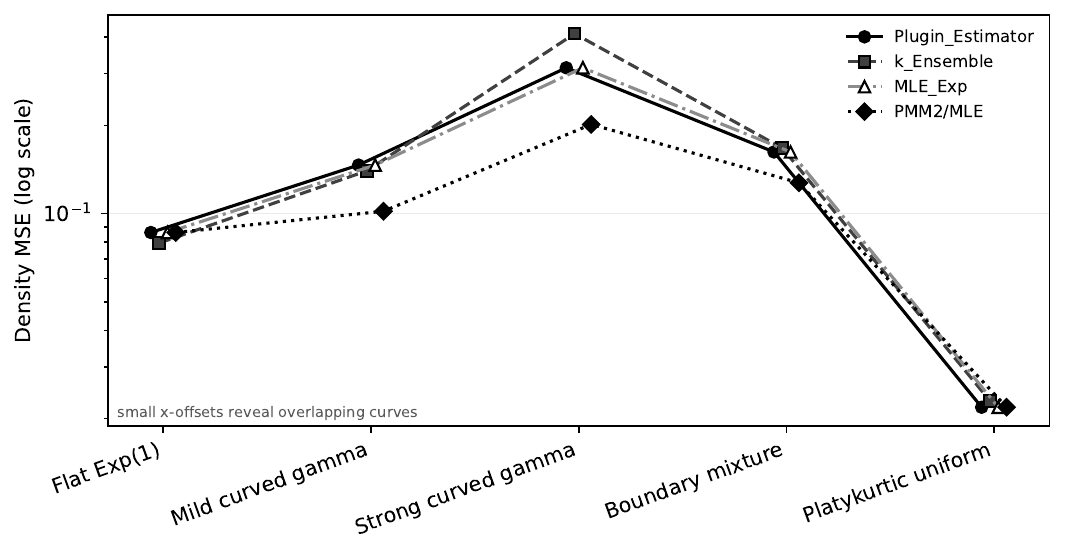}
\caption{Known-DGP density MSE by spacing regime.  PMM2 reduces MSE on
asymmetric gamma and boundary regimes while preserving the flat MLE fallback;
symmetric platykurtic regimes are separated into the PMM3-location diagnostic
study.  Small horizontal offsets are used only to reveal estimators whose
values overlap exactly.}
\label{fig:density_mse}
\end{figure}

\paragraph{PMM3 platykurtic diagnostic.}
The centered regression study follows the expected PMM3 pattern.  Near-normal
data have no practical PMM3 advantage: the PMM3/OLS variance ratios are
1.027, 1.054, and 1.035 for $n=50,100,200$.  As negative kurtosis increases,
the variance ratio decreases: triangular residuals give 0.988, 0.962, and
0.860; $\mathrm{Beta}(2,2)$ gives 0.796, 0.628, and 0.650; uniform residuals
give 0.446, 0.349, and 0.353; and strongly two-point-like residuals give
ratios below 0.01 for the larger sample sizes.  The positive-spacing
diagnostic is more restrictive but still informative.  With the guard active,
uniform spacings give PMM3-location density-MSE ratios of 0.482, 0.345,
0.317, and 0.302 for $k=16,32,64,128$, while $\mathrm{Beta}(2,2)$ spacings
give 0.746, 0.690, 0.671, and 0.660.  For the milder triangular law the guard
keeps the MLE ratio at 1.000.  These results support PMM3 only as a conditional
extension for strongly symmetric platykurtic spacing panels.

\paragraph{H2: resampling quality.}
The proxy layer does not establish end-to-end MASEM superiority.  It does show
that the density rule can matter operationally.  PMM improves the seven-lobes
proxy from $0.0074$ to $0.0044$ mean $W_2^2$ with Holm-adjusted
$p=0.001$, but it degrades the sine and swiss-roll proxies.  The swiss-roll
degradation is clearest: PMM increases proxy $W_2^2$ from $0.158$ to
$0.711$ with adjusted $p=0.001$.  These results support an applicability
boundary rather than a positive end-to-end result.

\input{tables/tab_e1.tex}

\begin{figure}[t]
\centering
\includegraphics[width=\linewidth]{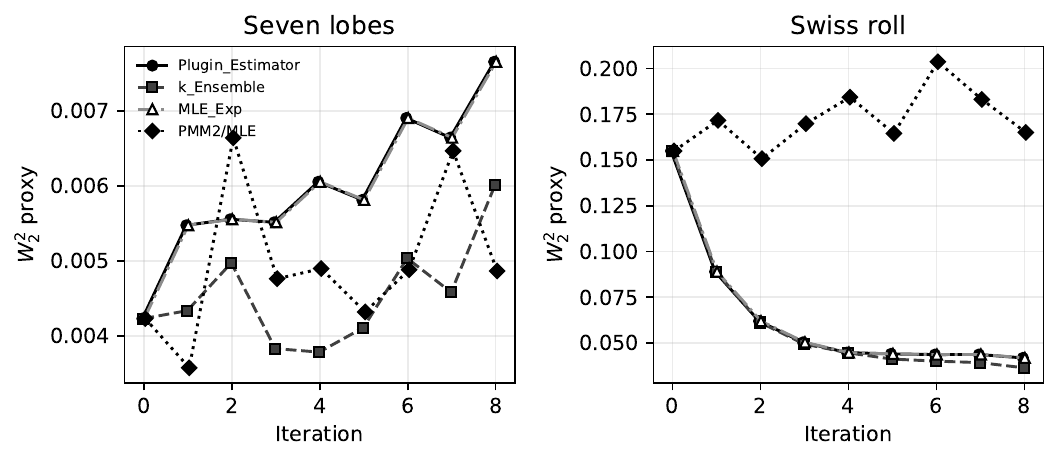}
\caption{Iterated resampling proxy for seven lobes and swiss roll.  The
matched-refresh proxy is diagnostic only; it does not replace a full MASEM
NHR/OLLA benchmark.  Small horizontal offsets reveal overlapping
Plugin/MLE curves.}
\label{fig:iteration_proxy}
\end{figure}

\paragraph{\texorpdfstring{$\tau$}{tau} tolerance.}
The proxy $\tau$ sweep is consistent with the mixed result.  On seven lobes,
PMM remains within 10\% of its $\tau=0.5$ value up to the largest tested
$\tau=1.5$, whereas Plugin and MLE tolerate only $\tau=0.5$ under the same
criterion.  On swiss roll and the robotics proxy, all estimators tolerate the
tested range, so this diagnostic does not distinguish them.

\input{tables/tab_e2.tex}

\begin{figure}[t]
\centering
\includegraphics[width=0.92\linewidth]{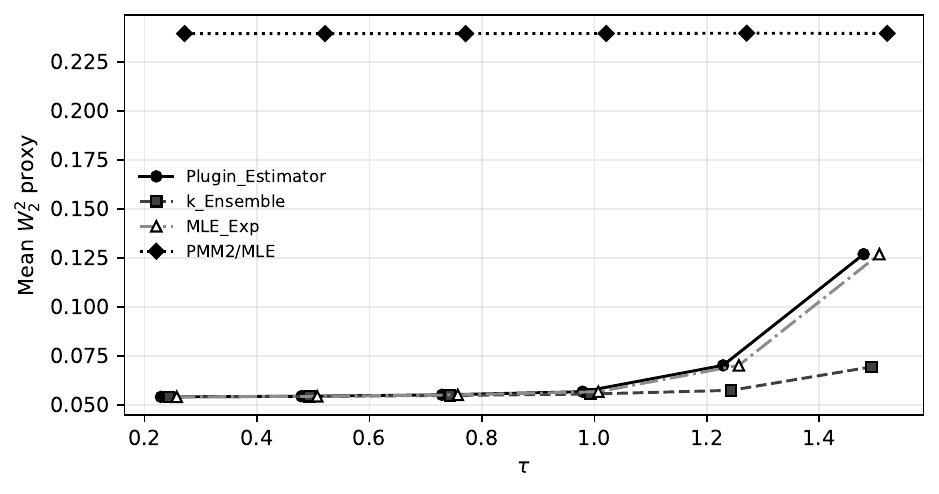}
\caption{Proxy $W_2^2$ as a function of resampling temperature $\tau$.
Small horizontal offsets reveal overlapping Plugin/MLE curves.}
\label{fig:tau_sweep}
\end{figure}

\paragraph{H3: component loss.}
Component coverage does not support a general PMM advantage.  In the component
sweep, Plugin, $k$-Ensemble, and MLE-Exp had zero component losses at all tested
particle budgets.  PMM2/MLE lost components in $6/15$ runs at $N=80$, $6/15$
runs at $N=140$, $0/15$ runs at $N=220$, and $1/15$ runs at $N=340$.  In the
main proxy table, sine and swiss-roll coverage also worsens under PMM.  H3 is
therefore not supported.

\paragraph{Computational overhead.}
In these small unjitted proxy runs, PMM is not the wall-clock bottleneck:
its mean estimator time is about 0.40 times the Plugin measurement, largely
because the Plugin path pays for the full $k$-NN radius computation while the
PMM/MLE paths reuse shell-spacing structure efficiently.  This estimate should
not be generalized to a compiled full MASEM run, but it indicates that the
PMM gate is not currently a dominant cost.

\begin{figure}[t]
\centering
\includegraphics[width=0.80\linewidth]{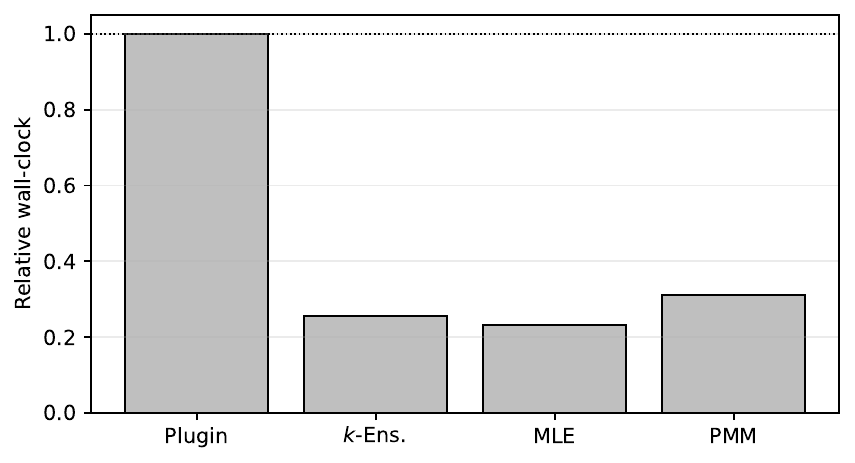}
\caption{Relative estimator wall-clock in the proxy experiment.}
\label{fig:wallclock}
\end{figure}

Overall, RQ1 is partially confirmed: PMM2 can reduce local density MSE under
asymmetric spacing misspecification while correctly falling back on flat
$\mathrm{Exp}(1)$, and PMM3-location has a narrower positive result under
strict symmetric platykurtic guards.  RQ2 is not supported as an end-to-end
MASEM claim:
the current proxy evidence is mixed and exposes failure modes on sine,
swiss-roll, and insufficiently guarded platykurtic regimes.

%% file: tables/tab_e1.tex
\begin{table*}[t]
\centering
\caption{Single-step resampling-proxy $W_2^2$ results.  Entries are mean $\pm$ 95\% CI over five seeds; parentheses report Holm-adjusted paired $p$-values against Plugin.  This table is weight-rule evidence, not a full NHR/OLLA MASEM benchmark.}
\label{tab:main_proxy_w2}
\resizebox{\textwidth}{!}{%
\begin{tabular}{lrrrr}
\toprule
Benchmark & Plugin & $k$-Ens. & MLE-Exp & PMM2/MLE \\
\midrule
Disconnected disks & 0.004 $\pm$ 0.000 & 0.004 $\pm$ 0.000 (0.886) & 0.004 $\pm$ 0.000 (--) & 0.003 $\pm$ 0.000 (0.032) \\
Seven lobes & 0.007 $\pm$ 0.002 & 0.005 $\pm$ 0.001 (0.004) & 0.007 $\pm$ 0.002 (--) & 0.004 $\pm$ 0.001 (0.001) \\
Sine curve & 4.905 $\pm$ 2.884 & 4.905 $\pm$ 2.884 (0.392) & 4.905 $\pm$ 2.884 (--) & 37.829 $\pm$ 18.495 (0.020) \\
Swiss roll & 0.158 $\pm$ 0.026 & 0.158 $\pm$ 0.026 (0.952) & 0.158 $\pm$ 0.026 (--) & 0.711 $\pm$ 0.141 (0.001) \\
Scaling stress-test & 0.013 $\pm$ 0.005 & 0.013 $\pm$ 0.005 (0.198) & 0.013 $\pm$ 0.005 (--) & 0.013 $\pm$ 0.005 (0.672) \\
Robotics corridor proxy & 0.003 $\pm$ 0.001 & 0.003 $\pm$ 0.001 (0.126) & 0.003 $\pm$ 0.001 (--) & 0.003 $\pm$ 0.001 (0.020) \\
\bottomrule
\end{tabular}%
}
\end{table*}

%% file: tables/tab_e2.tex
\begin{table}[t]
\centering
\caption{Proxy $\tau$ tolerance: largest tested $\tau$ whose mean $W_2^2$ stays within 10\% of the estimator's $\tau=0.50$ value.}
\label{tab:tau_tolerance}
\begin{tabular}{lrrrr}
\toprule
Benchmark & Plugin & $k$-Ens. & MLE-Exp & PMM2/MLE \\
\midrule
Seven lobes & 0.50 & 0.75 & 0.50 & 1.50 \\
Swiss roll & 1.50 & 1.50 & 1.50 & 1.50 \\
Robotics corridor proxy & 1.50 & 1.50 & 1.50 & 1.50 \\
\bottomrule
\end{tabular}
\end{table}

%% file: sections/07_discussion.tex
\section{Discussion}
\label{sec:discussion}

The present evidence validates \methodname{} as a local diagnostic and
density-estimation module rather than as a complete sampler improvement.  Its
main scientific value is a sharper statistical account of when the MASEM
density estimate has room for improvement.  In the flat homogeneous case, the
boundary is clear: the plug-in estimator is the MLE and should be retained.
Any useful PMM effect must come from curvature, boundary truncation, component
imbalance, or other misspecification of the Exp(1) spacing model.

This framing determines how negative results should be read.  If PMM improves the
known-DGP density task but not end-to-end sampling, then the local density
effect is real but not operationally decisive under the tested MASEM dynamics.
If PMM does not improve even the known-DGP task, then the PMM moment correction
is not appropriate for this spacing model as currently formulated.  If PMM
helps only under artificial curvature or boundary strength, then the result is
an applicability map rather than a general-purpose sampler improvement.

The evidence narrows the claim while strengthening its interpretation.  The
Known-DGP experiment justifies PMM2 as a variance-reduction candidate under
asymmetric spacing misspecification.  The PMM3 platykurtic study adds a second,
narrower message: symmetric negative-kurtosis structure can be exploitable, but
only when the estimator is posed as a guarded location equation for positive
spacings.  The sine and swiss-roll proxy failures show that these local gains
are insufficient to support a broad ``PMM improves MASEM'' conclusion.  The
main scientific message is therefore regime awareness: the kNN density
bottleneck is real, but moment corrections can be harmful when the spacing law
violates the branch assumptions.

The most important implementation limitation is cumulant estimation.  Global
pooling is stable but may blur component-specific geometry.  Local pooling is
more faithful but has higher variance.  Component-aware pooling after a coarse
partition may offer a useful compromise, but it would add another algorithmic
choice that must be validated.  PMM3 adds a second implementation constraint:
the centered regression score transfers to shell spacings only through a
location solve, positivity guard, and strict cumulant thresholds
($|c_3|\le0.3$, $c_4<-0.7$, $0<g_3<0.8$, and $k\ge16$ in the present study).
These conditions are natural for the platykurtic stress tests but should not
be assumed on arbitrary manifold clouds.

The theoretical limitation is the status of the curved-manifold cumulant
formulas.  The flat Exp(1) result is standard and strong.  The curvature
corrections are currently best treated as heuristic unless the derivation is
audited and matched by known-DGP simulations.  We therefore avoid theorem-level
claims based on these formulas.

Future work should first replace the proxy layer with the full MASEM
NHR/OLLA benchmark suite, then explore component-aware cumulant pooling,
adaptive temperature schedules, and alternative moment estimators for
boundary-heavy manifolds.

\S\ref{sec:conclusion} summarizes the contribution.

%% file: sections/08_conclusion.tex
\section{Conclusion}
\label{sec:conclusion}

This paper develops a PMM-gated density module for MASEM.  The module replaces
only the resampling density estimate, computes shell-spacing cumulants, falls
back to Plugin/MLE in the flat Exp(1) regime, activates PMM2 only under
validated asymmetric misspecification, and evaluates PMM3-location as a
guarded extension for symmetric platykurtic spacing panels.  The central
boundary is explicit: PMM should not be expected to outperform the plug-in
estimator on a flat homogeneous manifold because the plug-in rule is already
the MLE.

The current experiments support a boundary-paper conclusion.  PMM2 reduces
local density MSE on asymmetric gamma and boundary-spacing DGPs, PMM3-location
reduces mean and density MSE for strongly platykurtic positive spacings under
strict guards, and the resampling proxy is mixed: the seven-lobes proxy
improves, while sine and swiss roll degrade.  The contribution is therefore
not a finished general-purpose sampler improvement.  It is a validated
diagnostic and replacement module that identifies where PMM has statistical
room to help, where the MLE boundary should be retained, and where the present
selector remains unreliable.

%% file: appendix/appendix.tex
\section{Appendix Overview}
\label{app:overview}

This appendix records what must be included in the submission-ready version.
Detailed derivation drafts currently live in
\path{theory/flat_manifold.tex},
\path{theory/riemannian_manifold.tex}, and
\path{theory/quantitative_bridge.tex}.  They should be merged only after
the $g$/variance wording is audited.

\section{A.1 Flat-Manifold Spacing Derivation}
\label{app:flat}

For a locally homogeneous flat manifold, the count process in balls around
$x_i$ converges to a Poisson process in the volume coordinate
$u=N\rho(x_i)V_pr^p$.  Consecutive arrivals in that coordinate have i.i.d.
$\mathrm{Exp}(1)$ spacings.  Since
\[
  N\rho(x_i)\Delta_{i,j}
  =
  N\rho(x_i)V_p(\varepsilon_{i,j}^p-\varepsilon_{i,j-1}^p),
\]
the shell spacings inherit the Exp(1) law.  Summing spacings gives
$\sum_j\Delta_{i,j}=V_p\varepsilon_{i,k}^p$, hence the MLE is the plug-in
density estimator.

\section{A.2 Curved-Manifold Heuristic}
\label{app:curved}

The curved-manifold analysis starts from the geodesic ball expansion
\[
  \mathrm{vol}_g(B(x,r))
  =
  V_pr^p\left(1-\frac{\kappa r^2}{6(p+2)}+O(\kappa^2r^4)\right).
\]
This changes the spacing law observed by a density estimator that still uses
the Euclidean volume coordinate.  The final version must decide whether to
state the resulting cumulant shifts as theorem, approximation, or empirical
diagnostic.  Until Known-DGP MC confirms the formulas, this appendix should
use conservative language.

\section{A.3 PMM Implementation Details}
\label{app:pmm_implementation}

The implementation is vectorized in JAX.  Pairwise distances are computed for
all particles, sorted to obtain $k$-NN radii, and converted to shell spacings.
The selector computes all candidate densities and uses \texttt{jnp.where} for
traceable selection.  The flat gate tests proximity to $(c_3,c_4)=(2,6)$.
Invalid PMM denominators, negative variance coefficients, and boundary
cumulant regimes fall back to MLE.

\section{A.4 Full Tables}
\label{app:full_tables}

The generated main tables are included in the body as
Tab.~\ref{tab:spacing_regimes}, Tab.~\ref{tab:main_proxy_w2}, and
Tab.~\ref{tab:tau_tolerance}.  Their source CSVs are stored under
\texttt{results/}: \texttt{known\_dgp\_mc.csv},
\texttt{main\_benchmarks.csv}, and \texttt{tau\_tolerance.csv}.

\section{A.5 Known-DGP Monte Carlo Protocol}
\label{app:known_dgp}

Known-DGP experiments must include flat, curved, and boundary patches.  The
required outputs are density bias, density variance, MSE ratio relative to
Plugin/MLE, estimated $(c_3,c_4)$, and selector branch counts.  Flat patches
must trigger Plugin/MLE fallback.

\section{A.6 Reproducibility Manifest}
\label{app:reproducibility}

The current repository includes the local driver
\path{python3 -m experiments.run_all}, fixed seeds, generated result CSVs,
tables, and figures.  A public code supplement is available in the
\href{https://github.com/SZabolotnii/Ku-PMM-MASEM-code-supplement}{GitHub
repository}.  A DOI can be added after Zenodo archival release.

\section{A.7 AI-Use Disclosure}
\label{app:ai_use}

Drafting, code review, and planning assistance used an AI coding assistant.
All mathematical claims, experimental results, and final wording must be
reviewed and approved by the authors.  No result should be reported unless it
is generated by the repository scripts and traceable to saved artifacts.

%% file: refs.bib
@article{braun2026masem,
  title        = {Manifold Sampling via Entropy Maximization},
  author       = {Braun, C. V. and Burghoff, T. and Toussaint, M.},
  journal      = {arXiv preprint arXiv:2605.12338},
  eprint       = {2605.12338},
  archivePrefix = {arXiv},
  year         = {2026}
}

@article{loftsgaarden1965nonparametric,
  title        = {A Nonparametric Estimate of a Multivariate Density Function},
  author       = {Loftsgaarden, Don O. and Quesenberry, Charles P.},
  journal      = {The Annals of Mathematical Statistics},
  volume       = {36},
  number       = {3},
  pages        = {1049--1051},
  year         = {1965},
  doi          = {10.1214/aoms/1177700079}
}

@article{mack1979multivariate,
  title        = {Multivariate {$k$}-nearest neighbor density estimates},
  author       = {Mack, Y. P. and Rosenblatt, M.},
  journal      = {Journal of Multivariate Analysis},
  volume       = {9},
  number       = {1},
  pages        = {1--15},
  year         = {1979},
  doi          = {10.1016/0047-259X(79)90065-4}
}

@article{kozachenko1987sample,
  title        = {Sample Estimate of the Entropy of a Random Vector},
  author       = {Kozachenko, L. F. and Leonenko, N. N.},
  journal      = {Problems of Information Transmission},
  volume       = {23},
  number       = {2},
  pages        = {95--101},
  year         = {1987}
}

@article{pyke1965spacings,
  title        = {Spacings},
  author       = {Pyke, Ronald},
  journal      = {Journal of the Royal Statistical Society. Series B (Methodological)},
  volume       = {27},
  number       = {3},
  pages        = {395--436},
  year         = {1965},
  doi          = {10.1111/j.2517-6161.1965.tb00602.x}
}

@book{kunchenko1992stochastic,
  title        = {Stochastic Polynomials},
  author       = {Kunchenko, Yu. P.},
  publisher    = {Naukova Dumka},
  address      = {Kyiv},
  year         = {2006}
}

@incollection{warsza2018pmm,
  title        = {Estimation of Measurand Parameters for Data from Asymmetric Distributions by Polynomial Maximization Method},
  author       = {Warsza, Zygmunt Lech and Zabolotnii, Serhii V.},
  booktitle    = {Automation 2018},
  series       = {Advances in Intelligent Systems and Computing},
  volume       = {743},
  pages        = {746--757},
  publisher    = {Springer},
  address      = {Cham},
  year         = {2018},
  doi          = {10.1007/978-3-319-77179-3_74}
}

@article{zabolotnii2026estempmm,
  title         = {{EstemPMM}: Polynomial Maximization Method for Non-Gaussian Regression and Time Series in {R}},
  author        = {Zabolotnii, Serhii},
  journal       = {arXiv preprint arXiv:2605.02673},
  eprint        = {2605.02673},
  archivePrefix = {arXiv},
  primaryClass  = {stat.ME},
  year          = {2026},
  doi           = {10.48550/arXiv.2605.02673},
  url           = {https://arxiv.org/abs/2605.02673}
}

@article{zabolotnii2025arimapmm2,
  title         = {Applying the Polynomial Maximization Method to Estimate {ARIMA} Models with Asymmetric Non-{Gaussian} Innovations},
  author        = {Zabolotnii, Serhii},
  journal       = {arXiv preprint arXiv:2511.07059},
  eprint        = {2511.07059},
  archivePrefix = {arXiv},
  primaryClass  = {stat.ME},
  year          = {2025},
  doi           = {10.48550/arXiv.2511.07059},
  url           = {https://arxiv.org/abs/2511.07059}
}

@book{penrose2003random,
  title        = {Random Geometric Graphs},
  author       = {Penrose, Mathew},
  publisher    = {Oxford University Press},
  year         = {2003}
}

@book{kingman1993poisson,
  title        = {Poisson Processes},
  author       = {Kingman, J. F. C.},
  publisher    = {Oxford University Press},
  year         = {1993}
}

@article{pelletier2005kernel,
  title        = {Kernel Density Estimation on Riemannian Manifolds},
  author       = {Pelletier, Bruno},
  journal      = {Statistics \& Probability Letters},
  volume       = {73},
  number       = {3},
  pages        = {297--304},
  year         = {2005},
  doi          = {10.1016/j.spl.2005.04.004}
}

@article{henry2011nearest,
  title        = {{$k$}-Nearest Neighbor Density Estimation on Riemannian Manifolds},
  author       = {Henry, Guillermo and Mu{\~n}oz, Andr{\'e}s and Rodriguez, Daniela},
  journal      = {arXiv preprint arXiv:1106.4763},
  eprint       = {1106.4763},
  archivePrefix = {arXiv},
  year         = {2011}
}

@article{berry2017density,
  title        = {Density Estimation on Manifolds with Boundary},
  author       = {Berry, Tyrus and Sauer, Timothy},
  journal      = {Computational Statistics \& Data Analysis},
  volume       = {107},
  pages        = {1--17},
  year         = {2017},
  doi          = {10.1016/j.csda.2016.09.011}
}

@article{zhao2020analysis,
  title        = {Analysis of {KNN} Density Estimation},
  author       = {Zhao, Puning and Lai, Lifeng},
  journal      = {arXiv preprint arXiv:2010.00438},
  eprint       = {2010.00438},
  archivePrefix = {arXiv},
  year         = {2020}
}
